\documentstyle[12pt]{article}
\begin{document}

\title{First Passage Time in a Two-Layer System}
\author{Jysoo Lee and Joel Koplik \\
Benjamin Levich Institute and Department of Physics \\
City College of the City University of New York \\
New York, NY 10031}
\date{\today}
\maketitle

\begin{abstract}
As a first step in the first passage problem for passive tracer in
stratified porous media, we consider the case of a two-dimensional
system consisting of two layers with different convection velocities.
Using a lattice generating function formalism and a variety of
analytic and numerical techniques, we calculate the asymptotic
behavior of the first passage time probability distribution.  We show
analytically that the asymptotic distribution is a simple exponential in
time for any choice of the velocities. The decay constant is given in
terms of the largest eigenvalue of an operator related to a
half-space Green's function.  For the anti-symmetric case of opposite
velocities in the layers, we show that the decay constant for system
length $L$ crosses over from $L^{-2}$ behavior in diffusive limit to
$L^{-1}$ behavior in the convective regime, where the crossover length
$L^*$ is given in terms of the velocities. We also have formulated a
general self-consistency relation, from which we have developed a
recursive approach which is useful for studying the short time
behavior.

\end{abstract}


\newpage

\section{Introduction}
\label{sec:intro}

The motion of a passive tracer in a fluid under the combined action of
molecular diffusion and convection arises in a variety of settings,
such as fluid flows through porous media, fixed-bed catalytic
reactors, and the dispersion of pollutant in oceans \cite{gnp88}.  In
many situations, the convection-diffusion equation (CDE) describing
the variation of tracer concentration with space and time becomes
inhomogeneous, i.e., the fluid velocity field and/or the diffusivity
is not a constant, but a function of spatial position. One obvious
method in the study of inhomogeneous systems is a perturbation
technique \cite{d87,k87}. Here, one starts from a homogeneous version
of the system, which usually is solvable.  The velocity or diffusivity
is written as a sum of a homogeneous and an inhomogeneous term and the
appropriate quantities are expressed as expansions involving the
inhomogeneous term. The perturbative method is not immediately
applicable when the magnitude of the disorder is strong, as many or an
infinite number of terms are required, but in some cases, effective
medium theories may be used for an approximate summation
\cite{webman}. However such techniques are certainly not suitable for
systems whose disorder is strongly correlated in space, where usually
only straightforward numerical simulation or, for problems with an
appropriate geometry, network models \cite{sahimi}, are useful.

In this paper we are motivated by the particular case of tracer
dispersion in porous geological systems such as aquifers and
hydrocarbon reservoirs \cite{bear,dullien}, and by the observation
that such materials are very prominently stratified \cite{allen}. In
this context, Matheron and de Marsily \cite{mm80} first observed that
when the number of layers is effectively infinite, the velocity
fluctuations associated with the variation in structure and
permeability of the layers could give rise to superdiffusive tracer
motion.  Several authors studied this problem further
\cite{r89,bgkpr90,zkb90}, and by now there is a fair understanding of
the tracer probability distribution for the case of a large number of
horizontally-infinite layers.  Unfortunately, the results to date do
not provide concrete statements about the most practical
configuration, involving a source and sink of tracer at {\em finite}
separation.  One would like to solve the {\em first passage time}
problem for a large number of horizontal layers of finite extent, with
various boundary conditions (sink or reflection) at the system edges.
As a first step in this direction, we consider the simple case of
tracer motion in a geometry consisting of two two-dimensional,
semi-infinite layers, where tracer is released in the interior point
and is adsorbed at the edges. Although a great simplification compared
to the case of an infinite number of layers, as we shall see this
problem is already sufficiently difficult that only an approximate
solution is available.  (In fact, even in the ostensibly elementary
problem of simple diffusion in two half-spaces with different
diffusivities, a lengthy analysis has recently appeared \cite{wb}.)

The analysis to follow is based on an exact generating function
formalism for biased random walks in the geometry of interest, and
approximation schemes to extract the asymptotic behavior.  More
generally, we hope that our methods are pertinent to the problem of
transport in system with ``block'' disorder, for inhomogeneous
materials which are naturally modeled as a collection of finite
homogeneous sub-regions placed in contact \cite{ziman}.  When the size
of the sub-regions is much less than that of the system itself, or the
wavelength of any probe, the disorder is short ranged and perturbation
techniques are appropriate, but otherwise few methods beyond numerical
simulation are available.

In this paper, we address the first passage time properties of passive
tracer which convects and diffuses in a two layer system, which is the
simplest non-trivial case of layered structures. The system, shown in
Fig.~1, consists of two semi-infinite blocks occupying the
two-dimensional region $|x|\le L$.  The blocks are in physical
contact, allowing tracer to pass between them, and inside each block
different fluid flow fields convect the tracer. For simplicity, the
tracer diffusivities are assumed to be equal. The two finite
boundaries at $x = \pm L$ are taken to be perfect absorbers.  Tracer
is released at some point in the interior, and the time-dependent flux
at the boundary is computed, which in this situation is identical to
the first passage time probability distribution.

We begin in Section 2 with a precise formulation of the model as a
random walk process, and by introducing appropriate generating
functions, formulate an exact self-consistency relation for the first
passage time distribution. In Section 3, in order to obtain the
asymptotic behavior at long time, we expand the first passage time
distribution terms of the number of times a tracer particle has
crossed the interface between the blocks before reaching the
boundaries. We then approximately sum the expansion, using the central
limit theorem, to obtain the asymptotic distribution. We show
analytically that {\em the asymptotic distribution decays as a simple
exponential in time, for any choice of the velocity fields}, where the
decay constant is given in terms of the largest eigenvalue of an
operator which is related to a half-space Green's function.  We
estimate the decay constant for the special case of the
``antisymmetric'' model. For the limiting case of high velocities, we
estimate the largest eigenvalue and find the decay constant behaves as
$1/L$, which agrees with numerical simulations.  In the opposite case
of pure diffusion, the decay constant behaves as $1/L^2$, in good
agreement with analytic estimates and numerical simulations. In
Section 4, we consider the behavior in the intermediate velocity
regime, using two methods: an expansion method about the convective
limit, and a more general scaling argument which predicts a crossover
from a diffusive to a convective regime as $L$ increases.  The
crossover length $L^*$ is given in terms of the velocity, and the
scaling argument is consistent with the above results as well as those
of numerical simulations.  We conclude in Section 5, with a summary
and discussion of future possibilities. In Appendix A, we interpret
the general self-consistency condition as a recursion relation, and
obtain an expansion useful for obtaining the short time behavior of
the first pasage time distribution.  Appendix B gives solves the
first passage time problem explicitly for the simple case of convection
and diffusion in a single layer.

\section{Self-consistency Relation}
\label{sec:self}

\subsection{Definition of the Model}
\label{sec:def}

Since the tracer motion is given by the convection-diffusion equation,
one may equivalently think of it as a biased random walk on a spatial
lattice in discrete time.  Consider then a lattice of unit spacing in
the $x$-$y$ plane, where the velocity field takes on different values
in the upper and lower half-planes, and where only the region $-L \le
x \le L$ is relevant -- see Fig.~1.  The probability $P_n(x,y)$ that
the particle is at position $(x, y)$ at time $n$ is given by the
master equation
\begin{eqnarray}
\label{eq:master}
P_{n+1}(x,y) & = & p_x(y-1) ~p_y(y-1) ~P_n(x-1,y-1) \nonumber \\
	     & + & p_x(y+1) ~[1 - p_y(y+1)] ~P_n(x-1,y+1) \nonumber \\
             & + & [1 - p_x(y-1)] ~p_y(y-1) ~P_n(x+1,y-1) \nonumber \\
	     & + & [1 - p_x(y+1)] ~[1 - p_y(y+1)] ~P_n(x+1,y+1) \nonumber \\
	     & + & \delta _{n+1,0} ~\delta _{x,x_o} ~\delta _{y,y_o},
\end{eqnarray}
where $p_{x} ~(p_{y})$ are the hopping probabilities in the positive
$x~(y)$ direction, which satisfy
\begin{equation}
\label{eq:bias}
p_{x/y}(y) = \left\{ \begin{array}{ll}
                     p_{x/y}^u & \mbox{if y $\ge$ 1} \\
	             p_{x/y}^d & \mbox{otherwise},
                     \end{array}
\right.
\end{equation}
and where the Kroneker deltas prescribe that the particle starts from
$(x_o,y_o)$ at time $n=0$.  The master equation implies that each step
is along the diagonal of a square, which is a particularly convenient
hopping rule for the analysis to come, and in the limit of long time
and distance, as good as any other.  Indeed, by expanding the right
hand side in a Taylor series about $(x,y)$, it is easy to see that
(\ref{eq:master}) is equivalent to a convection-diffusion equation
with diffusion coefficient 1/2 and velocity $(2p_x-1,2p_y-1)$.  (There
are also higher order terms involving the derivatives of $p_{x/y}$
which are not relevant in the cases considered subsequently).

The first passage problem corresponds to absorbing boundaries at the
system edges, so we put $P_n(x,y)=0$ at $x =\pm L$, and define
$H_n^\pm$ to be the probability that the particle first reaches $x=\pm
L$ at time $n$.  Motivated by simplicity, and previous work on the
many-layer problem, we suppose the velocities are in the
$x$-direction, parallel to the layer boundaries, so that $p_{y}^u =
p_{y}^d = {1 \over 2}$.  For the same reasons, we assume that the net
convective bias or average velocity vanishes, which implies that the
probability of hopping to the right in the upper half plane equals the
probability of hopping to the left in the lower half plane, or
$p_{x}^u = 1 - p_{x}^d$.  We refer to this as the ``antisymmetric''
model, and some results about the general case appear in Section 5.
Lastly in the remainder of this section, we simplify the analysis by
further assuming $p_{x}^u = 1$ (and $p_{x}^d = 0$); so that particles
in the upper (lower) half plane always move to the right (left), and
we refer to this as the ``$+/-$ model.''  The latter restriction is
lifted in Section 4.

\subsection{Derivation of a self-consistency relation}

In this section, we derive some useful relations for the $+/-$ model.
The master equation (\ref{eq:master}) reduces to
\begin{eqnarray}
\label{eq:pmmaster}
P_{n+1}(x,y) & = & \left\{\begin{array}{ll}
               ( P_n(x-1,y-1) + P_n(x-1,y+1) ) / 2 & y \ge 2 \\
	       ( P_n(x-1,y+1) + P_n(x+1,y-1) ) / 2 & y = 0, 1 \\
               ( P_n(x+1,y-1) + P_n(x+1,y+1) ) / 2 & y \le -1
		          \end{array} \right. \nonumber \\
	     & + & \delta_{n+1,0} ~\delta_{x,x_o} ~\delta_{y,y_o}.
\end{eqnarray}
We define the following generating functions
\begin{eqnarray}
\label{eq:generate}
P(x,y,z) & \equiv & \sum_{n=0}^{\infty} P_n(x,y) z^n, \nonumber \\
G_+(x,\alpha,z) & \equiv & \sum_{y=1}^{\infty} P(x,y,z) \alpha^y,
\nonumber \\
G_-(x,\alpha,z) & \equiv & \sum_{y=-\infty}^{0} P(x,y,z) \alpha^y.
\end{eqnarray}
Substituting Eq.~(\ref{eq:generate}) in Eq.~(\ref{eq:pmmaster}), we
obtain
\begin{eqnarray}
G_+(x,\alpha,z) & = & {z \over 2} (\alpha + {1 \over \alpha})
G_+(x-1,\alpha,z) \nonumber \\
                & + & {z \over 2} (P(x+1,0,z)\alpha - P(x-1,1,z)) \nonumber \\
		& + & \delta_{x,x_o} \alpha^{y_o}
\end{eqnarray}
and
\begin{eqnarray}
G_-(x,\alpha,z) & = & {z \over 2} (\alpha + {1 \over \alpha})
G_-(x+1,\alpha,z) \nonumber \\ & + & {z \over 2} (P(x-1,1,z) -
P(x+1,0,z)\alpha ),
\end{eqnarray}
where we assume $y_o \ge 1$ without the loss of generality.

The functions $G_+$ and $G_-$ can be expressed in terms of simple
Green's functions. We define the Green's function in the upper block
($y \ge 1$) to be the solution of
\begin{equation}
g_+(x;x',\alpha,z) = {z \over 2}(\alpha + {1 \over \alpha})
g_+(x-1;x',\alpha,z) + \delta_{x,x'},
\end{equation}
which is
\begin{equation}
g_+(x;x',\alpha,z) = \left\{ \begin{array}{ll}
({z \over 2}(\alpha + {1 \over \alpha}))^{x-x'} & \mbox{if} ~-L < x'
\le x < L \\ 0 & \mbox{otherwise.}
\end{array} \right.
\end{equation}
Similarly, the Green's function for the lower block ($y \le 0$), the
solution of
\begin{equation}
g_-(x;x',\alpha,z) = {z \over 2}(\alpha + {1 \over \alpha})
g_-(x+1;x',\alpha,z) + \delta_{x,x'},
\end{equation}
which is
\begin{equation}
g_-(x;x',\alpha,z) = \left\{ \begin{array}{ll}
({z \over 2}(\alpha + {1 \over \alpha}))^{x'-x} & \mbox{if} ~-L < x
\le x' < L \\ 0 & \mbox{otherwise.}
\end{array} \right.
\end{equation}
Using these Green's functions, $G_{\pm}$ can be expressed as
\begin{eqnarray}
\label{eq:green+}
G_+(x,\alpha,z) & = & {z \over 2} \sum_{x' = -L + 1}^{L - 1}
g_+(x;x',\alpha,z) (P(x'+1,0,z)\alpha - P(x'-1,1,z)) \nonumber \\
		& + & g_+(x;x_o,\alpha,z) \alpha^{y_o},
\end{eqnarray}
and
\begin{equation}
\label{eq:green-}
G_-(x,\alpha,z) = {z \over 2} \sum_{x' = -L + 1}^{L - 1}
g_-(x;x',\alpha,z) (P(x'-1,1,z) - P(x'+1,0,z)\alpha )
\end{equation}
There is a simple way to understand Eqs.~(\ref{eq:green+}) -
(\ref{eq:green-}). The Green's function $g_+ ~(g_-)$ is the solution
of the homogeneous convection equation $p_x = 1 ~(0)$ for a particle
starting from $(x',0)$. The second term in Eq.~(\ref{eq:green+})
corresponds to the original particle which starts from $(x_o,y_o)$.
The first term of the equation is due to the existence of the
boundary. It subtracts the contribution of the particle [$P(x,1,z)$
term] which leaves the block, and add the contribution of the particle
[$P(x,0,z)$ term] which enters the block. The equation for the lower
block Eq.~(\ref{eq:green-}) has essentially the same structure, except
it lacks the second term due to the absence of a starting particle in
the block.

We are interested in the first passage properties which can be
calculated from $H_n^{\pm}$. They are related to $G_{\pm}$ by
\begin{eqnarray}
\label{eq:hit}
H^+(z) & = & z ~G_+(L-1,1,z) \nonumber \\
H^-(z) & = & z ~G_-(-L+1,1,z),
\end{eqnarray}
where $H^{\pm}(z)$ is defined to be $\sum_{n=0}^{\infty} H_n^{\pm}
z^{n}$. In Eqs.~(\ref{eq:green+}) - (\ref{eq:green-}), $G_{\pm}$ are
expressed in terms of two unknown functions $P(x,0,z)$ and $P(x,1,z)$,
which again can be calculated from $G_{\pm}$ themselves as follows. We
expand Eq.~(\ref{eq:green+}) as a series of $\alpha$. The terms
proportional to $\alpha$, from the definition of $G_+(x,\alpha,z)$,
are exactly $P(x,1,z) ~\alpha$. In other words,
\begin{eqnarray}
\label{eq:self+}
P(x,1,z) & = & {z \over 2} \sum_{x'=-L+1}^x ({z \over 2})^{x-x'}
{x - x' \choose (x - x') / 2} ~P(x'+1,0,z) \nonumber \\
	 & - & {z \over 2} \sum_{x'=-L+1}^x ({z \over 2})^{x-x'}
{x - x' \choose (x - x' + 1) / 2} ~P(x'-1,1,z) \nonumber \\
	 & + & ({z \over 2})^{x-x_o} ~{x - x_o \choose (x - x_o + y_o -1)/2}
\end{eqnarray}
Similarly, $P(x,0,z)$ can be expressed as
\begin{eqnarray}
\label{eq:self-}
P(x,0,z) & = & {z \over 2} \sum_{x'=x}^{L-1} ({z \over 2})^{x'-x}
{x' - x \choose (x' - x) / 2} ~P(x'-1,1,z) \nonumber \\
	 & - & {z \over 2} \sum_{x'=x}^{L-1} ({z \over 2})^{x'-x}
{x' - x \choose (x' - x - 1) / 2} ~P(x'+1,0,z),
\end{eqnarray}
where ${x \choose y}$ is the binomial coefficient. Here, we also
define ${x \choose y} = 0$, if $x$ or $y$ is not a non-negative
integer, or if $x < y$. Thus, the problem of calculating the first
passage property is reduced to solving the self-consistency equations
Eqs.~(\ref{eq:self+}) - (\ref{eq:self-}). This is the key result in
this section, and it will later serve as a basis for an iteration
scheme.  It is not unnatural that we end up with self-consistent
relations rather than explicit solutions.  The boundary conditions we
have to satisfy at the interface between the two blocks are (1)
continuity and (2) flux conservation. Since these conditions are only
implicit [i.e., they are relations among the fields $P(x,y,z)$], they
result in implicit relations between $G(x,\alpha,z)$, which are the
self-consistency conditions.

Unfortunately, these conditions are essentially $4L-2$ coupled linear
equations, which are non-trivial to solve. We have developed an
iterative scheme useful in getting the short time behaviors ($n \sim
L$), which is discussed in Appendix A. We now develop an alternative
method which can give the information about the asymptotic ($n \gg L$)
behavior.

\section{Asymptotic Behavior}
\label{sec:asym}

\subsection{Expansion of $H_+(z)$}

We turn to an alternative method for obtaining the first passage time.
We expand quantities in terms of the number of times the particle has
crossed the interface between the blocks before reaching the boundary.
Consider the $+/-$ model again Eq.~(\ref{eq:pmmaster}), and recall the
previous definitions of $P(x,y,z)$ and $H_+(z)$
(Eqs.~(\ref{eq:generate}) and (\ref{eq:hit})).  We now define
``half-space'' Green's functions $g_{\pm}^h$ as
\begin{eqnarray}
\label{eq:hgreen}
g_+^h(x,y,x',y',z) & = & ({z \over 2})^{x-x'} \left\{ {x-x' \choose
(x-x'+y-y')/2} - {x-x' \choose (x-x'+y+y')/2} \right\} \nonumber \\
g_-^h(x,y,x',y',z) & = & ({z \over 2})^{x'-x} \left\{ {x'-x \choose
(x'-x+y-y')/2} - {x'-x \choose (x'-x+y+y'-2)/2} \right\}, \nonumber \\
		   &   & \nonumber \\
		   &   &
\end{eqnarray}
where ${x \choose y}$ is defined as in Eq.~(\ref{eq:self-}). Here,
$g_+^h$ ($g_-^h$) is the Green's function in the upper (lower) block
with absorbing boundary at $y = 0$ ($y = 1$). Due to the boundary
condition, the functions $g_{\pm}^h$ do not contain contributions from
the particles which leave the block, a property which will be useful
subsequently.

We define $H_+^{(n)}(z)$, the part of $H_+(z)$ corresponding to
particles which have crossed the interface $n$ times. Using the
definition of $g_+^h$,
\begin{equation}
\label{eq:h0}
H_+^{(0)}(z) = \sum_y z g_+^h(L-1,y,x_o,y_o,z),
\end{equation}
where we assume $y_o > 0$ without the loss of generality. We now
calculate the flux of particles out of the upper block. Define
$P^{(n)}(x,y,z)$ to be the part of $P(x,y,z)$ corresponding to
particles which have crossed the interface $n$ times. At the edge of
the upper block (the $y = 1$ line),
\begin{equation}
P^{(0)}(x,1,z) = g_+^h(x,1,x_o,y_o,z).
\end{equation}
Half ($1 - p_y^u$) of these particle will jump to $(x + 1, 0)$.
Therefore, the influx at the edge of the lower block ($y = 0$ line),
\begin{equation}
\label{eq:flux-}
P^{(1)}_{\mbox{in}}(x,0,z) \equiv {z \over 2} P^{(0)}(x-1,1,z).
\end{equation}
We define the operator $T_{\pm}(x,x',z)$ as
\begin{equation}
{z \over 2} v(x \pm 1,z) \equiv \sum_{x'} T_{\pm}(x,x',z) v(x,z'),
\end{equation}
where $v(x,z)$ is a vector. Thus, Eq.~(\ref{eq:flux-}) in operator
form is
\begin{equation}
P^{(1)}_{\mbox{in}}(0,z) = T_-(z) P^{(0)}(1,z),
\end{equation}
where we dropped the indices $x$ and $x'$.

We have to know what fraction the flux will go back to the upper
block. We first obtain
\begin{equation}
P^{(1)}(x,0,z) = \sum_{x'} g_-^h(x,0,x',0,z) P^{(1)}_{\mbox{in}}
(x',0,z).
\end{equation}
Again, half ($p_y^d$) of the particles will cross the interface and
jump to $(x-1 ,1)$,
\begin{equation}
P^{(2)}_{\mbox{in}}(x,1,z) \equiv {z \over 2} P^{(1)}(x+1,0,z) =
T_+(z) P^{(1)}(0,z).
\end{equation}
Since $H_+^{(2)}$ arises from from the walkers which have crossed the
interface twice, its sole contribution comes from
$P^{(2)}_{\mbox{in}}$, which is
\begin{equation}
H_+^{(2)}(z) = z \sum_y \sum_{x'} g_+^h(L-1,y,x',1,z) P^{(2)}
_{\mbox{in}} (x',1,z).
\end{equation}
We then calculate the fraction of $P^{(2)}_{\mbox{in}}$ which jumps
back to the lower block, thus completing the cycle. At the edge of the
upper block,
\begin{equation}
P^{(2)}(x,1,z) = \sum_{x'} g_+^h(x,1,x',1,z) P^{(2)}_{\mbox{in}}(x,1,z),
\end{equation}
and half of these will jump to $(x+1,0)$
\begin{equation}
P^{(3)}_{\mbox{in}}(0,z) = T_-(z) P^{(2)}(1,z).
\end{equation}
The above results can be written in a more compact form. We first
define several operators
\begin{eqnarray}
\label{eq:fdef}
(g_+^o(z))_{x,x'} & \equiv & g_+^h(x,1,x',1,z), \nonumber \\
(g_-^o(z))_{x,x'} & \equiv & g_-^h(x,0,x',0,z), \nonumber \\
(h_+^o(z))_{x,x'} & \equiv & z \sum_y g_+^h(L-1,y,x',1,z).
\end{eqnarray}
in terms of which the above results can be written as
\begin{eqnarray}
P^{(2)}_{\mbox{in}}(1,z) & = & T_+(z) g_-^o(z) T_-(z) P^{(0)}(1,z),
\nonumber \\
H_+^{(2)}(z) & = & h_+^o(z) P^{(2)}_{\mbox{in}}(1,z).
\end{eqnarray}
Furthermore, by repeating the above procedure, we can show that
\begin{eqnarray}
P^{(2i)}_{\mbox{in}}(1,z) & = & T_+(z) g_-^o(z) T_-(z) g_+^o(z)
P^{(2i-2)}_{\mbox{in}}(1,z), \nonumber \\
H_+^{(2i)}(z) & = & h_+^o(z) P^{(2i)}_{\mbox{in}}(1,z).
\end{eqnarray}
With one more definition
\begin{eqnarray}
u(z) & = & T_+(z) g_-^o(z) T_-(z) g_+^o(z), \nonumber \\
u_1(z) & = & T_+(z) g_-^o(z) T_-(z),
\end{eqnarray}
we arrive to the key result of this section:
\begin{equation}
\label{eq:hsum}
H_+(z) = H^{(0)}(z) + \sum_{i=0}^{\infty} h_+^o(z) u^i(z) u_1(z)
P^{(0)}(1,z).
\end{equation}
The validity of the expansion has been checked by comparing $H_+(z)$
obtained above with that obtained by numerical simulations. Details of
the simulations will be discussed later.

\subsection{Asymptotic form of $H_+^n$}

In this section, we derive the asymptotic form of the first passage
time distribution, starting from Eq.~(\ref{eq:hsum}). Recall the
definition $H_+(z) \equiv \sum_{n} H_+^n z^n$. In general, $H_+(z)$ is
an infinite order polynomial of $z$, where $H_+^n$, the coefficient of
$z^n$, is the hitting probability at time $n$. We now consider the
various terms in Eq.~(\ref{eq:hsum}). Using Eq.~(\ref{eq:h0}), we can
show the degree of $H_+^{(0)}(z)$ can not be larger than $2L -1$.
Since it does not give a contribution to $H_+(z)$ in the asymptotic $n
\gg L$ regime, we can ignore this term. Next, in the summand of the
equation, the same operator $u(z)$ is been repeatedly applied to a
vector $u'(z) P^{(0)}(1,z)$. Thus, we can approximate $u^i(z)$ with
$\lambda ^i(z)$, where $\lambda (z)$ is the largest eigenvalue of
$u(z)$ If the operator in question is self-adjoint and diagonalizable,
this approximation would surely be justified, at least for $i \gg
1/(\lambda (z) - \lambda _2 (z))$, where $\lambda _2(z)$ is the second
largest eigenvalue of $u(z)$, but in this instance this step is an
assumption, which however is supported by the numerical results below.
We now ask whether the asymptotic behavior will be changed by the
approximation. The maximum degree of $z$ for the terms in the summand
can be calculated from Eq.~(\ref{eq:hgreen}).  The maximum degree of
$h_+^o(z), u(z), u'(z)$ and $P^{(0)}(1,z)$ cannot be larger than
$2L-1, 4L-2, 2L, 2L-2,$ respectively. In the sum, the term containing
$u^i(z)$ contributes for $n \le i (2L-1) + 6L -3$.  Therefore, the
terms for which the eigenvalue approximation is not valid ($i < i_o$,
and $i_o$ is finite) give a contribution only up to finite time, and
will not change the asymptotic behavior. Thus,
\begin{equation}
H_+(z) \sim \sum_{i=0}^{\infty} \lambda ^i (z) \cdot [h_+^o(z) u'(z)
P^{(0)}(1,z)].
\end{equation}
The product $h_+^o(z) u'(z) P^{(0)}(1,z)$ is also a polynomial of
finite order, and ignoring the product changes only the amplitude of
the asymptotic behavior. We thus arrive to a simple expression
\begin{equation}
\label{eq:lsum}
H_+(z) \sim {1 \over 1 - \lambda (z)} = \sum_{i=0}^{\infty} \lambda
^i (z).
\end{equation}

We define the coefficient of the $z^{j}$ term of $\lambda (z)$ to be
$c_j^1$.  We can interpret $\lambda (z)$ as a generating function for
a random walk process---the probability to jump $j$ steps forward is
given by the coefficient $c_j^1$. The fraction of random walkers which
survive after one step is $s_o \equiv \lambda (1)$. The average
displacement after one step is $s_1 \equiv \lambda '(1) / \lambda
(1)$, and the average of the square of the displacement after one step
is $s_2
\equiv \lambda ''(1) / \lambda (1)$, where
\begin{eqnarray}
\label{eq:pdef}
\lambda' (z) & \equiv & z{d \over dz} \lambda (z), \nonumber \\
\lambda'' (z) & \equiv & [z{d \over dz}]^2 \lambda (z).
\end{eqnarray}
We also define the variance $\sigma ^2 \equiv s_1^2 - s_2 $.
Following the interpretation, the coefficient $c_j^i$ of the $z^j$
term for $\lambda ^i (z)$ forms the distribution of the displacement
of the random walker after $i$ steps. The fraction of random walkers
which survive for $i$ steps is $s_o^i$, the average displacement is $i
s_1$, and variance $i \sigma ^ 2$. Since the second moment $s_2$ is
finite, we can apply the central limit theorem. Thus for large $i$,
the coefficient $c_j^i$ becomes
\begin{equation}
c_j^i \sim {s_o^i \over \sqrt{2 \pi i \sigma ^2}} \exp[-{(j - i s_1)^2
\over 2 i \sigma ^2}].
\end{equation}
Substituting this to the equation for $H_+(z)$ Eq.~(\ref{eq:lsum}), we
obtain
\begin{equation}
H_+^n \sim \sum_{i=0}^{\infty} {s_o^i \over \sqrt{2 \pi i \sigma ^2}}
\exp[-{(n - i s_1)^2 \over 2 i \sigma ^2}],
\end{equation}
which can be evaluated by the method of the steepest descent to be
\begin{eqnarray}
\label{eq:hasum}
H_+^n & \sim & \exp [\ln s_o \cdot {n \over s_1} \cdot {1 - \ln s_o
\cdot (\sigma / s_1)^2 / 2 \over 1 - \ln s_o \cdot (\sigma / s_1)^2}]
\nonumber \\
      & \equiv & \exp[-c(L) n],
\end{eqnarray}
where $c(L)$ is size dependent decay constant. Even though the
equation is derived for the $+/-$ model, its derivation only assumes
the existence of the half space Green's functions similar to
Eq.~(\ref{eq:hgreen}). {\em Since these functions exist for the most
general situation of the two block system, the asymptotic distribution
is a always simple exponential}. Below, we compare the above results
with an exact enumeration method, and find good agreement.

\subsection{Estimation of the Eigenvalue}
\label{sec:estl}

The problem of finding the asymptotic behavior of the first passage
time distribution is reduced to finding $\lambda (z),$ the largest
eigenvalue of the operator $u(z)$. Unfortunately, there is no known
method to calculate the analytic expression of the eigenvalues of an
arbitrary matrix, and the complicated structure of $u(z)$ does not
help matters. We present two methods to estimate $\lambda (z)$. These
methods are not expected to produce exact numbers, but are intended to
give some idea of the parameter $L$-dependences of the first passage
time distribution.

The first method is to express $\lambda (z)$ in terms of the average
of the elements of $u(z)$. We start from the matrix $T_+(z) g_-^o(z)$,
whose largest eigenvalue $\lambda _+(z)$ is approximated as
\begin{equation}
\label{eq:eigen}
	\lambda _+ (z) \sim {1 \over 2L - 1} \sum_{x=-L+1}^{L-1}
	\sum_{x'=-L+1}^{L-1} (T_+(z) g_-^o(z) )_{x,x'},
\end{equation}
where $2L - 1$ is the size of the matrix. This approximation is
motivated from the numerical fact that the eigenvector $v_+(z)$
corresponding to $\lambda _+ (z)$ is close to be uniform, i.e.,
$(v_+(z))_x = v_o(z)$ for all $x$. Note that if the eigenvector is
uniform, Eq.~(\ref{eq:eigen}) becomes exact. We then obtain
\begin{eqnarray}
\lambda _+(z) & \sim & {1 \over 2L-1} ({z \over 2}) (2L-2) \nonumber \\
              & + & {1 \over 2L-1} \sum_{k=1}^{L-1} ({z \over 2})^{2k+1}
              (2L - 2k -2) \left\{ {2k \choose k} - {2k \choose k+1}
	          \right\}.
\end{eqnarray}
The expression can be further simplified to
\begin{equation}
\label{eq:lplus1}
\lambda _+(z) \sim {z \over 2} + {1 \over \sqrt{4 \pi}} \sum_{k=1}^L
z^{2k} k^{-3/2} \left(1 - {k \over L} \right),
\end{equation}
where we assume $L \gg 1$. The largest eigenvalue $\lambda _-(z)$ of
the matrix $T_-(z) g_+^o(z)$ can also be estimated by the same method.
It turns out
\begin{equation}
\lambda _-(z) \sim \lambda _+(z).
\end{equation}
The matrix $u(z)$ is given by the product of the two matrices, $T_+(z)
g_-^o(z)$ and $T_-(z) g_+^o(z)$. We further approximate the largest
eigenvalue of the product of two matrices as product of the largest
eigenvalues of the two matrices, which implies
\begin{equation}
\label{eq:lp2l}
\lambda (z) \sim \lambda _-(z) \lambda _+(z) \sim \lambda _+^2(z).
\end{equation}
We calculate $\lambda _+(1), \lambda _+'(1)$ and $\lambda _+''(1)$,
where the primed values are defined as the same way in
Eq.~(\ref{eq:pdef}). By evaluating the integral in
Eq.~(\ref{eq:lplus1}), we obtain
\begin{eqnarray}
\label{eq:l1}
\lambda _+(1) & \sim & \left({1 \over 2} + {1 \over \sqrt{\pi}}\right)
              - {2 \over \sqrt{\pi}} L^{-1/2} + {\cal O}(1/L),
	      \nonumber \\
\lambda _+'(1) & \sim & {4 \over 3 \sqrt{\pi}} L^{1/2} +\left({1 \over
	      2} - {2 \over \sqrt{\pi}} \right) + {\cal O}(1/L),
	      \nonumber \\
\lambda _+''(1) & \sim & {8 \over 15 \sqrt{\pi}} L^{3/2} + \left( {1
	      \over 2} - {4 \over 3 \sqrt{\pi}} \right) + {\cal O}(1/L).
\end{eqnarray}

The second approximation method is based on the interpretation that
$\lambda (z)$ is related to a certain generating function for a random
walk. Consider the matrix $T_-(1) g_+^o(1)$. The matrix gives the
probability to reach points on $y=0$, starting from points on $y=1$
with an absorbing boundary at $y=0$. Thus, $\lambda _+(z)$ is roughly
the generating function of the hitting probability on the line $y=0$
for a walk starting from $y=1$. For simplicity, consider a walker
starting from $(0,1)$. Since the only effect of convection in the $x$
direction is to remove all the walkers which do not reach $y=0$ until
time step $L$, we only have to deal with an one-dimensional problem.
The corresponding one-dimensional problem is treated by ignoring the
$x$ axis and limiting the maximum time step to be $L$. Then $\lambda
_+(z)$ is
\begin{equation}
\lambda _+(z) \sim \int_{1}^{L} dn {z^n \over \sqrt{2 \pi n}} (1 -
\exp^{-8/n}),
\end{equation}
where the integrand is the flux to $y=0$ at step $n$, and we have
approximated the sum by an integral. The eigenvalue $\lambda _+(1)$ is
\begin{equation}
\lambda _+(1) \sim \int_{1}^{\infty} dn {1 \over \sqrt{2 \pi n}} (1
- \exp^{-8/n}) - \int_{L}^{\infty} dn {1 \over \sqrt{2 \pi n}} (1 -
\exp^{-8/n}).
\end{equation}
The first integral is the probability to hit $y=0$ during infinite
period of time, which is unity. After simplifying the second integral
for $L \gg 1$ we have
\begin{equation}
\lambda _+(1) \sim 1 - {16 \over \sqrt{2 \pi}} L^{-1/2}.
\end{equation}
The second approximation, compared to Eq.~(\ref{eq:l1}), has the same
dependence on $L$, but different numerical coefficients. This supports
the previous suggestion that the coefficients obtained by these
methods are not reliable. However, the fact that two very different
methods give the same dependence on $L$ gives some support to the
validity of the form. The leading term in $\lambda _+(1)$ , which is
the value of $\lambda _+(1)$ in the limit $L \to\infty$, deserves
special attention.  It is the probability that an unbiased random
walker hits the $y=0$ line during an infinite period of time, which is
equal to unity.  Even in the case that the matrix is applied to the
exact eigenvector, the random walker eventually has to be absorbed at
the $y=0$ boundary for $L \to \infty$, implying $\lambda _+(1) = 1$.
Therefore, we set $\lim _{L \to \infty} \lambda _+(1) = 1$ from now
on.

We now proceed to calculate $c(L)$. The expressions for $s_o, s_1$ and
$s_2$ can be obtained from Eqs.~(\ref{eq:l1}) and (\ref{eq:lp2l}):
\begin{eqnarray}
\label{eq:lest}
s_o & \sim & \lambda _+^2(1) \simeq 1 - {4 \over \sqrt{\pi}} L^{-1/2}
 + {\cal O}(1/L), \nonumber \\
s_1 & \sim & 2 ~{\lambda '_+(1) \over \lambda _+(1)} \simeq {8 \over
3 \sqrt{\pi}} L^{1/2} + {\cal O}(1), \nonumber \\
s_2 & \sim & 2 ~{\lambda ''_+(1) \over \lambda _+(1)} + 2 \left(
{\lambda '_+(1) \over \lambda '_+(1)} \right)^2 \simeq {16 \over 15
\sqrt{\pi}} L^{3/2} + {\cal O}(L), \nonumber \\
\sigma ^2 & \sim & s_2 - s_1^2 \sim {16 \over 15 \sqrt{\pi}}
L^{3/2} + {\cal O}(L).
\end{eqnarray}
Finally, we can calculate $c(L)$ using the definition in
Eq.~(\ref{eq:hasum}),
\begin{eqnarray}
\label{eq:dcon}
c(L) & \equiv & -{\ln s_o \over s_1} ~{1 - \ln s_o (\sigma / s_1)^2
/ 2 \over 1 - \ln s_o (\sigma / s_1)^2} \nonumber \\
& \sim & {39 \over 32} L^{-1}.
\end{eqnarray}

\subsection{Numerical Check}

We have calculated the decay constant $c(L)$ in three major steps.
First, we expand $H_+(z)$ in terms of the number of times a random
walker have crossed the interface (Eq.~(\ref{eq:hsum})). We then
calculate its asymptotic distribution in terms of the eigenvalue
$\lambda (z)$ with the help of the central limit theorem
(Eq.~(\ref{eq:hasum})). We then estimate $\lambda (z)$ and $c(L)$
(Eqs.~(\ref{eq:lest}) - (\ref{eq:dcon})). In this section, we check
the validity of these results by comparing them with those of
numerical simulations. It will serve as an intermediate check before
we proceed to more general situation, where we will continue to use
the methods developed above.

We start with the exact sum Eq.~(\ref{eq:hsum}). We calculate first
few moments of $H_+(z)$ from the equation, which are easier to
compare. The zero-th moment $H_+(1)$ is rather easy to evaluate.
Consider the $i$-th term in the sum. We just have to calculate the
product $h_+^o(1) u(1) P_{\mbox{in}}^{(2i+2)}(1,1),$ where we know all
the individual terms. Furthermore, $(i+1)$-th term can be obtained by
replacing $P_{\mbox{in}}^{(2i+2)}(1,1)$ by $P_{\mbox{in}}
^{(2i+4)}(1,1) = u(1) P_{\mbox{in}} ^{(2i+2)}(1,1)$. We sum these
terms in the increasing order of $i$, until the magnitude of the new
term is smaller than certain value, which is chosen to be $10^{-20}$.
The higher moments are slightly more complicated to calculate.
Consider again $i$-th term in the sum. The first moment can be
calculated by using the chain rule
\begin{eqnarray}
[ h_+^o(z) u(z) P_{\mbox{in}}^{(2i+2)}(1,z) ]^{\prime}
& = & h_+^{\prime o}(z) u(z) P_{\mbox{in}}^{(2i+2)}(1,z)
+ h_+^o(z) u^{\prime} (z) P_{\mbox{in}}^{(2i+2)}(1,z)
\nonumber \\
& + & h_+^o(z) u(z) P_{\mbox{in}}^{\prime (2i+2)}(1,z),
\end{eqnarray}
where the primed values are defined as the same way in
Eq.~(\ref{eq:pdef}). Also $(i+1)$-th term can be obtained by replacing
\begin{eqnarray}
P_{\mbox{in}}^{(2i+2)}(1,z) & \rightarrow & P_{\mbox{in}}^{(2i+4)}
(1,z) = u(z) P_{\mbox{in}}^{(2i+2)}(1,z), \nonumber \\
P_{\mbox{in}}^{\prime (2i+2)}(1,z) & \rightarrow & P_{\mbox{in}}^
{\prime (2i+4)}(1,z) = u^{\prime} (z) P_{\mbox{in}}^{(2i+2)}(1,z)
+ u(z) P_{\mbox{in}}^{\prime (2i+2)}(1,z). \nonumber \\
& &
\end{eqnarray}
Higher order moments are calculated by same way with the heavy use of
the identity
\begin{equation}
[A(z) B(z)]^{(n)} = \sum_{k=0}^{n} {n \choose k} A^{(k)}(z)
B^{(n-k)}(z),
\end{equation}
where $A^{(n)}(z)$ is the $n$-th derivative of the function $A(z)$.
We compare the first five moments with those obtained by an exact
enumeration method \cite{ha87}. For several value of $L=10 \sim 100$
and several initial conditions, the values obtained by the two methods
are essentially identical.

We check the next step, the asymptotic form of the hitting probability
distribution Eq.~(\ref{eq:hasum}). The form is simple exponential, and
the decay constant $c(L)$ is given in terms of $\lambda (z)$. We
directly calculate the distribution $H_+^n$ by the exact enumeration
method for $L=10$ to $300$. In Fig.~2, we show the distribution for $L
= 100$. It is clear that $H_+^n$ is an exponential after some
transition period. This is also true for the other sizes we have
studied, and the length of the transition period is of the order $L$.

We check the value of the decay constant $c(L)$.  Since the
theoretical value of $c(L)$ is given in terms of $\lambda (z)$, we
have to know the value of $\lambda (z)$ in order to compare. To
calculate the eigenvalue numerically, we go back to the discussion in
the previous paragraph about calculating the moments of $H_+(z)$. We
consider $\lambda (1)$ first. Since the matrix $u(1)$ has been
repeatly applied to the vector $P_{\mbox{in}}^{(2i+2)}(1,1)$ to obtain
$P_{\mbox{in}}^{(2i+4)}(1,1)$,
\begin{equation}
\lambda (1) = \lim _{i \to \infty} P_{\mbox{in}}^{(2i+4)}(1,1) /
P_{\mbox{in}}^{(2i+2)}(1,1).
\end{equation}
We find the ratio hardly changes for $i > 50$, so we take the ratio at
$i = 100$ as $\lambda (1)$. For higher moments, start from the
relation,
\begin{equation}
\lambda (z) = \lim _{i \to \infty} P_{\mbox{in}}^{(2i+4)}(1,z)
/ P_{\mbox{in}}^{(2i+2)}(1,z).
\end{equation}
Taking the derivative and multiply $z$ on both sides
\begin{equation}
\lambda ^{\prime} (1) = \lim _{i \to \infty}
{P_{\mbox{in}}^{\prime (2i+4)}(1,1) P_{\mbox{in}}^{(2i+2)}(1,1)
-P_{\mbox{in}}^{(2i+4)}(1,1) P_{\mbox{in}}^{\prime (2i+2)}(1,1)
\over (P_{\mbox{in}}^{(2i+2)}(1,1))^2}.
\end{equation}
The higher order terms (e.g., $\lambda ^{\prime \prime}(1)$) can be
calculated in a similar way. Finally, using these $\lambda (1),
\lambda ^{\prime}(1)$ and $\lambda ^{\prime \prime}(1)$, the decay
constant $c(L)$ is calculated from Eq.~(\ref{eq:hasum}). In Fig.~3, we
show the values of $c(L)$ just obtained as well as those obtained by
numerical simulations, for several values of $L$. The simulational
values are obtained by least square fitting the last $1/2$ or $1/3$
part of the numerically obtained $H_+^n$, like the one in Fig.~2.
There are in general good agreements between these two values except
for very small values of $L$, where several approximations made to get
the theoretical value may not be justified.

We proceed to the last step of the calculation, the estimation of the
decay constant $c(L)$. In Fig.~4, we show the values of $c(L)$ given
by Eq.~(\ref{eq:dcon}) and those obtained by the exact enumeration.
The values by the enumeration are identical to those shown in Fig.~3.
It is clear that the numerical data shows the $1/L$ behavior as
predicted by the theory. On the other hand, the measured prefactor
($\sim 1$) is little smaller than the predicted value ($39/32$). These
are all in accord to the expectation that the prediction of the $1/L$
dependence is reliable, but that of the prefactor is not. It is
unexpected that the value of prefactor by the enumeration is so close
to that of the theory.

We have checked the steps to reach the decay constant $c(L)$. The
errors involved in the eigenvalue approximation are well controlled,
and the approximation seems to be well justified for obtaining the
asymptotic properties. Even though we do not have the same level of
rigor in estimating the eigenvalues $\lambda (z)$, we still have
enough control to predict the correct dependence of $L$.

\section{The General Antisymmetric Model}
\label{sec:sym}

Having obtained a reasonable understanding of the asymptotic behavior
of the $+/-$ model, we turn to the general antisymmetric model. We now
allow an arbitrary horizontal bias $0\le p_x^u\le 1$, and due to the
symmetry in the system, we can restrict ourselves to $p_x^u \ge 1/2$
without loss of generality.

\subsection{The Diffusive Limit: $p_x^u = 1/2$}

We consider the antisymmetric model with no bias ($p_x^u = 1/2$),
first using the formalism developed for the $+/-$ model. We now have a
different operator $u(z)$, and therefore a different eigenvalue
$\lambda (z)$. The asymptotic distribution of $H_+^n$ is still a
simple exponential, and the decay constant $c(L,p_x^u)$ is given in
terms of $\lambda (z)$ (Eq.~(\ref{eq:dcon})).  In Sec.~\ref{sec:estl}
one estimate was based on by transforming the problem into an one
dimensional diffusion problem with an absorbing boundary.  In the
transformation, one determines the average time required for the
particles to be absorbed at the external boundaries at $x = \pm L$. In
the $+/-$ model, the transport in the horizontal direction is a purely
convective process, so that the time is identical to the length $L$.
Now, the transport in the horizontal direction is purely diffusive, so
the time is $2 L^2$. Substituting into Eq.~(\ref{eq:dcon}),
\begin{equation}
\label{eq:nbdcon}
c(L,p_x^u) \sim {29 \over 64} L^{-2},
\end{equation}
The $L^{-2}$ dependence is also consistent with the calculation for
the one block case with no convection. In Fig.~5, we plot $c(L,p_x^u)$
determined by the above equation as well as those determined by the
exact enumeration. The numerical data clearly shows the $1/L^2$
dependence with the prefactor about twice of that given in
Eq.~(\ref{eq:nbdcon}).

A more direct check of these results is obtained by noting that in
this case we are considering pure diffusion in a two dimensional strip
of width $2L$, and one expects an exponential decay of tracer
concentration with a time constant $O(L^2)$.  More precisely, a
straightforward solution of the diffusion equation for this geometry
in Appendix B  gives $H_n^+\sim \exp[-(\pi^2/8L^2)n]$, in good
agreement with the above simulation.

\subsection{The Neighborhood of $p_x^u = 1$}

It is useful to explicitly consider the case $p_x^u = 1 - \epsilon$ to
first order of $\epsilon$ and the formalism developed for the $+/-$
model can be used with only minor changes.  Starting from the master
equation (\ref{eq:master}), it is straightforward to show that the
half-space Green's functions $g_+^h, g_-^h$ Eq.~(\ref{eq:hgreen}) must
be modified to
\begin{eqnarray}
g_+^h(x,y;x',y',z)
& = & (1-(x-x')\epsilon)({z \over 2})^{x-x'} \nonumber \\
& \times & \left\{{x-x' \choose (x-x'+y-y')/2}
               -  {x-x' \choose (x-x'+y+y')/2} \right\} \nonumber \\
& + & (x-x'+2) \epsilon ({z \over 2})^{x-x'+2}  \nonumber \\
& \times & \left\{{x-x'+2 \choose (x-x'+y-y'+2)/2}
               -  {x-x'+2 \choose (x-x'+y+y'+2)/2} \right\} \nonumber \\
g_-^h(x,y;x',y',z)
& = & (1-(x'-x)\epsilon)({z \over 2})^{x'-x} \nonumber \\
& \times & \left\{{x'-x \choose (x'-x+y-y')/2}
               -  {x'-x \choose (x'-x+y+y'-2)/2} \right\} \nonumber \\
& + & (x'-x+2) \epsilon ({z \over 2})^{x'-x+2} \nonumber \\
& \times & \left\{{x'-x+2 \choose (x'-x+y-y'+2)/2}
               -  {x'-x+2 \choose (x'-x+y+y')/2}
\right\}, \nonumber \\
& & \nonumber \\
& &
\end{eqnarray}
where terms of order $\epsilon^2$ are ignored. The above equation
reduces to Eq.~(\ref{eq:hgreen}) for $\epsilon = 0$, which suggests
that the perturbation is not singular. The eigenvalue $\lambda (z)$
for the matrix $u(z)$ can be obtained by following the same procedure
as in the $+/-$ model. We first calculate the eigenvalue
$\lambda_+(z)$ of the matrix $T_-(z) g_+^o(z)$
\begin{eqnarray}
\lambda _+(z) & \sim & \lambda _+(z) |_{\epsilon = 0} \nonumber \\
              & + & {2 \epsilon \over 2L-1} \sum_{k=1}^{L-1} (2k+1)
	      ({z \over 2})^{2k+1} \left\{ {2k \choose k} - {2k
	      \choose k-1} \right\},
\end{eqnarray}
where $\lambda _+(z)|_{\epsilon = 0}$ is the value of $\lambda _+(z)$
at $\epsilon = 0$, and $g_+^o(z)$ is defined as in
Eq.~(\ref{eq:fdef}).  Replacing the sum by an integral, we obtain
\begin{eqnarray}
\lambda _+(1) & \sim & 1 - {2 \over \sqrt{\pi}} (1 - \epsilon) L^{-1/2}
+ {\cal O}(1/L) \nonumber \\
\lambda _+'(1) & \sim & {4 \over 3 \sqrt{\pi}} (1 + \epsilon) L^{1/2}
+ {\cal O}(1)
\nonumber \\
\lambda _+''(1) & \sim & {8 \over 15 \sqrt{\pi}} (1 + 3 \epsilon)
L^{3/2} + {\cal O}(L^{1/2}),
\end{eqnarray}
where we have set $\lim_{L \to \infty} \lambda _+(1) = 1$ as discussed
above. Similarly, the eigenvalue $\lambda _-(z)$ of the matrix
$T_+(z) g_-^o(z)$ is determined to be
\begin{equation}
\lambda _-(z) \sim \lambda _+(z).
\end{equation}
Combining these relations, the eigenvalue $\lambda (z)$ can be written as
\begin{equation}
\lambda (z) \sim \lambda _-(z) \lambda _+(z) \sim \lambda _+^2(z).
\end{equation}
 From $\lambda (z)$, the decay constant $c(L,p_x^u)$ is determined to
be
\begin{equation}
\label{eq:expan}
c(L,p_x^u) \sim {39 \over 32} ~(1 - 2 \epsilon) L^{-1}.
\end{equation}
which is the same as the $\epsilon = 0$ result aside from the factor
$1 - 2 \epsilon$, whose origin is easy to understand. When we estimate
the eigenvalue by mapping into an one dimensional problem, the average
absorption time is required. For pure convection, this time is the
length $L$ divided by the horizontal velocity.  In general, this
velocity is given by $2 p_x^u - 1=1 -2 \epsilon$. Therefore, in the
absence of diffusion, $c(L,p_x^u)$ has to be modified by replacing $L$
by $L / (1-2\epsilon),$ which is precisely what we find by the
expansion. In Fig.~6, we show $c(L,p_x^u)$ for $L=100$ and several
values of $p_x^u$, where $c(L,p_x^u)$ is divided by the value at
$p_x^u = 1$. We also show the corresponding result by the expansion
Eq.~(\ref{eq:expan}). There is a good agreement between the two, and
the expansion is not only valid near $p_x^u=1$ but there is no
systematic deviation down to $p_x^u = 0.65$.

\subsection{Scaling Argument for General $p_x^u$}

The half space Green's functions Eq.~(\ref{eq:hgreen}) have served as
a starting point in the calculation of the decay constant at $p_x^u =
1$ and its neighborhood. For other values of $p_x^u$, unfortunately,
we are unable to find them in closed form. Although we can still show
that the asymptotic distribution of $H_+^n$ is a simple exponential,
without explicit knowledge of the Green's functions, we no longer can
use the same method to estimate the eigenvalue and the decay constant.

However, since we know the behavior for the two extreme cases of the
antisymmetric model ($p_x^u = 1/2$ and $1$), we can try to bridge the
gap by a simple scaling argument. We propose a scaling {\em ansatz}
\begin{equation}
c(L,p_x^u) = L^{-2} ~f ( {L \over L^*} ),
\end{equation}
where $f(x)$ is a scaling function which satisfies
\begin{equation}
\label{eq:scale}
f(x) \sim \left\{ \begin{array}{ll}
	          1   & \mbox{if} ~x \ll 1 \\
	          x   & \mbox{if} ~x \gg 1.
                  \end{array} \right.
\end{equation}
where we have defined a crossover length $L^*=1/(2p_x^u-1)$. (Recall
that the lattice spacing has been set to one.)  To verify that the
{\em ansatz} is in accord with previous results, note first that for
$p_x^u = 1$, $L^*=1$, and since we are interested in $L\gg 1$,
(\ref{eq:scale}) gives $c(L,p_x^u) \sim L^{-1}$ as before.  Next, for
$p_x^u = 1/2$, $L^*\to\infty$ and $c(L,p_x^u) \sim L^{-2}$, again in
agreement.  Finally, for $p_x^u = 1 - \epsilon$, $L^*=1/(1 - 2
\epsilon)$, and for small $\epsilon$, $L / L^*\gg 1$, so the decay
constant becomes $(1-2\epsilon)/L$, which is exactly
Eq.~(\ref{eq:expan}). To verify the {\em ansatz} away from the
limiting cases, we have used numerical simulation.  In Fig.~7, we show
the rescaled $c(L,p_x^u)$ obtained by exact enumeration {\em vs.} the
rescaled $L$, for several values of $L=10 \sim 200$ and $p_x^u = 0.5
\sim 1.0$. The data collapse into one scaling curve, which approaches
a constant as $x \to 0$ and is linear for large $x$, precisely as
expected from Eq.~(\ref{eq:scale}).  Thus the scaling {\em ansatz}
provides an excellent description of the general antisymmetric model.

\section{Conclusions}
\label{sec:con}

We have studied the first passage time distribution $H_+^n$ of a two
layer system of width $L$, and determined its asymptotic form to be an
simple exponential decay in time. For the special case of an
antisymmetric model, the decay constant is calculated using several
techniques, and is found to cross over from the expected $L^{-2}$
behavior in the pure-diffusion regime to an $L^{-1}$ behavior at high
velocities.

The origin of the $L^{-1}$ behavior in the convective regime is not
intuitively obvious to us.  It arises as the result of two
contributions---one $L^{-1/2}$ factor from $\ln s_o$ term, and another
$L^{-1/2}$ factor from the $1 / s_1$ term. As discussed in
Sec.~\ref{sec:asym}, $s_o$ is roughly an eventual absorption
probability of a one dimensional random walk, and $s_1$ is mean
distance traveled before the absorption. This differs from a naive
expectation that the $L^{-1}$ behavior results from the mean distance
$s_1$ behaving as $L$ in the convective regime.

Evidently, we have only considered the more tractable special cases in
a two-layer system. It would be desirable to go beyond the
antisymmetric limit of zero average velocity. In terms of the $p_x^u -
p_x^d$ plane, the antisymmetric model corresponds to the line $p_x^u =
1 - p_x^d$, and the (elementary) one block case to the line $p_x^u =
p_x^d$.  Due to the symmetries, the remaining region is bounded by the
two cases with $1/2 \le p_x^u \le 1$. Of course, one would like to
consider convection in two directions as well multiple layers, but
these rather more difficult problems must await further work.

\section*{Acknowledgements}

This research was supported in part by the Department of Energy under
grant DE-FG02-93-ER14327.

\section*{Appendix A -- Iterative Method for Short Time Behavior}

We discuss an iterative scheme to obtain an approximate
solution of the self-consistency equations Eqs.~(\ref{eq:self+}) -
(\ref{eq:self-}), based on interpreting them
as a recursion relation. We input an trial solution
of $P(x,0,z)$ and $P(x,1,z)$ to the equations, and obtain a
(hopefully) improved approximation. In principle, we repeat this
procedure, until it converges to the correct solution.

We start from a trial solution
\begin{eqnarray}
P^{(0)}(x,0,z) & = & 0, \nonumber \\
P^{(0)}(x,1,z) & = & 0,
\end{eqnarray}
where the superscript indicates the number of iterations. For
simplicity, we set $x_o = 0$ and $y_o = 1$. As shown in
Eq.~(\ref{eq:hit}), the hitting probability $H_+$ is related to $G_+$
via $H_+(z) = z G_+(L-1,1,z)$. Also, using Eq.~(\ref{eq:green+}),
$G_+(L-1,1,z)$ can be written as
\begin{eqnarray}
G_+(L-1,1,z) & = & {1 \over 2} \sum_{x'=-L+1}^{L-1} z^{L-x'} (P(x'+1,0,z) -
P(x'-1,1,z)) \nonumber \\
             & + & z^{L-1}.
\end{eqnarray}
Combining these relations, we obtain
\begin{equation}
H_+^{(0)}(z) = z^{L}.
\end{equation}
This rather trivial result is due to the fact that the presence of the
second block is ignored in the $0^{\rm th}$ order approximation.

The next order values of $P(x,0,z)$ and $P(x,1,z)$ can be calculated
by inserting the trial values into Eqs.~(\ref{eq:self+}) -
(\ref{eq:self-}) to obtain
\begin{eqnarray}
P^{(1)}(x,0,z) & = & 0 \nonumber \\
P^{(1)}(x,1,z) & = & ({z \over 2})^x {x \choose x/2},
\end{eqnarray}
where we define ${x \choose y} \equiv 0$, if $x$ or $y$ is not a
non-negative integer, or if $x < y$. Similarly, we obtain
\begin{equation}
H_+^{(1)}(z) = z^L - {1 \over 2} z^L \sum_{x'=0}^{L-2} ({1 \over
2})^{x'} {x' \choose x'/2}.
\end{equation}
Using Sterling's formula, and replacing the sum by an integral,
\begin{equation}
H_+^{(1)}(z) \simeq z^L ( 1 - \sqrt{2 \over \pi} L^{1/2} ).
\end{equation}
The above result is unphysical, since $H_+$ becomes negative for large
$L$. The deficiency is due to the fact that we only include
the flux {\em out} of
the upper block and not the flux {\em into} the block, while both
fluxes are of the same order of magnitude. This problem will be
resolved in the calculation at next order. The $2^{\rm nd}$ order
iterations of $P(x,0,z)$ and $P(x,1,z)$ are
\begin{eqnarray}
P^{(2)}(x,0,z) & = & {z \over 2} \sum_{x'=\mbox{max}(x,1)}^{L-1} ({z \over
2})^{x'-x} {x'-x \choose (x'-x)/2} ({z \over 2})^{x'-1} {x' - 1 \choose
(x'-1)/2} \nonumber \\
P^{(2)}(x,1,z) & = & ({z \over 2})^{x} {x \choose x/2} \nonumber \\
               & - & {z \over 2} \sum_{x'=1}^x ({z \over 2})^{x-x'}
               {x-x' \choose (x-x'+1)/2} ({z \over 2})^{x'-1} {x' - 1
               \choose (x' -1 )/2}.
\end{eqnarray}
Again, using Sterling's formula and replacing the sum by an integral.
\begin{eqnarray}
P^{(2)}(x,0,z) & = & {1 \over \pi} \int_{\mbox{max}(x,1)}^{L-1} dx'
z^{2x' - x} {1 \over \sqrt{(x' - x)(x'-1)}} \nonumber \\
P^{(2)}(x,1,z) & = & z^x \sqrt{2 \over \pi x} - {z^x \over \pi}
	             \int_1^x dx' {1 \over \sqrt{(x - x')(x'-1)}},
\end{eqnarray}
and $H_+^{(2)}$ becomes
\begin{eqnarray}
H_+^{(2)}(z) & = & z^L \nonumber \\
	     & - & {z^L \over \sqrt{2 \pi}} \int_1^{L-1} dx {1 \over
                    \sqrt{x - 1}} \nonumber \\
	     & + & {z^L \over 2 \pi} \int_2^{L-1} dx \int_1^{x-1} dx'
                   {1 \over \sqrt{(x-1-x')(x'-1)}} \nonumber \\
             & + & {z^L \over 2 \pi} \int_{-L+1}^0 dx \int_1^{L-1} dx'
                   z^{2x'-2x} {1 \over \sqrt{(x'-x-1)(x-1)}} \nonumber \\
             & + & {z^L \over 2 \pi} \int_1^{L-1} dx \int_{x+1}^{L-1} dx'
                   z^{2x'-2x} {1 \over \sqrt{(x'-x-1)(x-1)}}.
\end{eqnarray}
Although the integrals in the equation can not be evaluated in closed
form, we can understand their structure. The first term is the
contribution of tracer which did not cross the boundary. The second
and the third terms are amount of flux going out of the first block.
Thus, these three terms are proportional to $z^L$. The fourth and
fifth terms are contributions from walkers which return to the upper
block. Since the time to reach the boundary ($x=L$) depends on the
where the walker exits ($x'$) and reenters ($x$) the upper block,
these terms contain different orders of $z$.

\section*{Appendix B -- The Single-layer System}

To provide some feeling for the more difficult case of a non-zero
average velocity, we calculate the first passage time distribution for
a single layer.  Consider tracer moving between adsorbing boundaries
at $x=\pm L$ in the presence of a constant velocity field $v\hat{x}$.
The motion in the $y$-direction is simple diffusion, completely decoupled
from that along $x$, and the problem effectively is one-dimensional.
We have
\begin{equation}
{\partial c\over\partial t}+v{\partial c\over\partial x}=
D{\partial^2 c\over\partial x} \quad\mbox{with}\quad c(\pm L,t)=0,
\end{equation}
with the simple initial condition $c(x,0)=\delta (x)$.  Taking the Laplace
transform via $\int_0^\infty dt\ e^{-st}$, we have
\begin{equation}
s c-\delta (x)+v{\partial c\over\partial x}=
D{\partial^2 c\over\partial x},
\end{equation}
which is readily solved for $x\ne 0$ as
\begin{equation}
c^{\pm}(x,s)=A^{\pm}
e^{px/L}\sinh{\left[ \sqrt{p^2+\sigma} (1\mp {x\over L})\right] }.
\end{equation}
Here the superscripts refer to $x>0$ and $x<0$, respectively, and we
have defined $p=vL/2D$ and $\sigma =sL^2/D$.  The coefficients $A^\pm$
are determined by the conditions $c^+=c^-$ and $\partial c^+ /\partial
x-\partial c^-/\partial x =-1/D$ at $x=0$, which follow from the
differential equation, so that
\begin{equation}
A^+=A^- ={ L \over 2D\sqrt{p^2+\sigma}\cosh{\sqrt{p^2+\sigma}} }.
\end{equation}
The Laplace transform of the flux leaving the system at $x=\pm L$,
which is identical to the first passage time probability distribution,
is
\begin{equation}
J^{\pm}(s)=-D{\partial c(\pm L,s)\over\partial x}
={e^{\pm p}\over 2\cosh{\sqrt{p^2+\sigma}} }.
\end{equation}
The long-time asymptotic behavior of $J^{\pm}$ is controlled by the
right-most singularities of the Laplace transform in the complex-$s$
plane, in this case the poles where $\sqrt{p^2+\sigma}=\pm i\pi /2$ or
$s=s^*= -\pi^2D/4L^2-v^2/4D$.  Thus, for $t\sim\infty$,
$J^{\pm}(t)\sim e^{s^*t}$.  In the pure-diffusion limit, we set $v=0$
and recall that $D=1/2$ and identify $t$ with step number $n$, so that
$H^+_n=J^+(n)\sim e^{-\pi^2n/8L^2}$.  In the opposite limit of large
velocity, we see that $J^+(t)\sim e^{-v^2t/4D}$, which coincides with
the long-time behavior at a fixed spatial point of the usual Gaussian
solution of the CDE.

\newpage
\section*{Figure Captions}

\begin{description}

\item [Fig.~1:] System geometry:  two semi-infinite blocks --
$y > 0$ and $y \le 0$ -- with different velocities, with
absorbing boundaries at $x = \pm L$.

\item [Fig.~2:] The hitting probability distribution $H_+^n$ for
$L=100$ obtained by the exact enumeration. An exponential behavior in
the asymptotic regime is evident.

\item [Fig.~3:] The decay constant $c(L)$ for the $+/-$ model given
by the eigenvalue approximation (solid line) compared to the values
obtained by the exact enumeration (diamonds).

\item [Fig.~4:] The decay constants $c(L)$ for the $+/-$ model given
by Eq.~(\ref{eq:dcon}) (dashed line) and those obtained by
the exact enumeration (diamonds).

\item [Fig.~5:] The decay constants $c(L,p_x^u)$ for the antisymmetric
model with $p_x^u = 1/2$ given by Eq.~(\ref{eq:dcon})
(solid line) and those obtained by the exact enumeration (diamonds).

\item [Fig.~6:] The normalized decay constants $c(L,p_x^u)$ for the
symmetric model predicted by the theory (solid line), and data
obtained by the enumeration for $L=100$ and several values of $p_x^u$
(diamonds). There is good agreement even down to $p_x^u = 0.65$.

\item [Fig.~7:] The rescaled decay constants $c(L,p_x^u)$ obtained by
the exact enumeration {\em vs.} rescaled $L$. Data for several different
values of $L = 10 \sim 200$ and $p_x^u = 0.5 \sim 1.0$ collapse into
one curve, which is the scaling function $f(x)$.

\end{description}

\end{document}